\documentclass[conference,11pt]{IEEEtran}

\usepackage[hyphens]{url} % For the problem of the URLs are too long in References
\usepackage[hidelinks]{hyperref} % For the problem of the URLs are too long in References
\usepackage{multirow}   % \multirow のため
\usepackage{array}
\usepackage{enumitem}

\usepackage{float}  % プリアンブルに追加

\usepackage{graphicx}

\title{Network Intrusion Detection: Evolution from Conventional Approaches to LLM Collaboration and Emerging Risks}

\author{
    \IEEEauthorblockN{Yaokai Feng}
    \IEEEauthorblockA{
        Faculty of Information Science and Electrical \\
        Engineering, Kyushu University, Japan\\
        Email: fengyk@ait.kyushu-u.ac.jp
    }
    \and
    \IEEEauthorblockN{Kouichi Sakurai}
    \IEEEauthorblockA{
         Faculty of Information Science and Electrical\\
         Engineering, Kyushu University, Japan\\      
        Email: sakurai@inf.kyushu-u.ac.jp
    }
}

\begin{document}

\maketitle

\thispagestyle{plain} % add pagenumber also on the first page

\begin{abstract}
This survey systematizes the evolution of network intrusion detection systems (NIDS), from conventional methods such as signature-based and neural network (NN)-based approaches to recent integrations with large language models (LLMs). It clearly and concisely summarizes the current status, strengths, and limitations of conventional techniques, and explores the practical benefits of integrating LLMs into NIDS.
Recent research on the application of LLMs to NIDS in diverse environments is reviewed, including conventional network infrastructures, autonomous vehicle environments and IoT environments.

From this survey, readers will learn that: 1) the earliest methods, signature-based NIDSs, continue to make significant contributions to modern systems, despite their well-known weaknesses; 2) NN-based detection, although considered promising and under development for more than two decades, and despite numerous related approaches, still faces significant challenges in practical deployment;
3) LLMs are useful for NIDS in many cases, and a number of related approaches have been proposed; however, they still face significant challenges in practical applications. Moreover, they can even be exploited as offensive tools, such as for generating malware, crafting phishing messages, or launching cyberattacks. Recently, several studies have been proposed to address these challenges, which are also reviewed in this survey; and
4) strategies for constructing domain-specific LLMs have been proposed and are outlined in this survey, as it is nearly impossible to train a NIDS-specific LLM from scratch.

\end{abstract}

\section{Organization of This Survey Paper}

\begin{figure}[htbp]
  \centering
  \includegraphics[width=\linewidth]{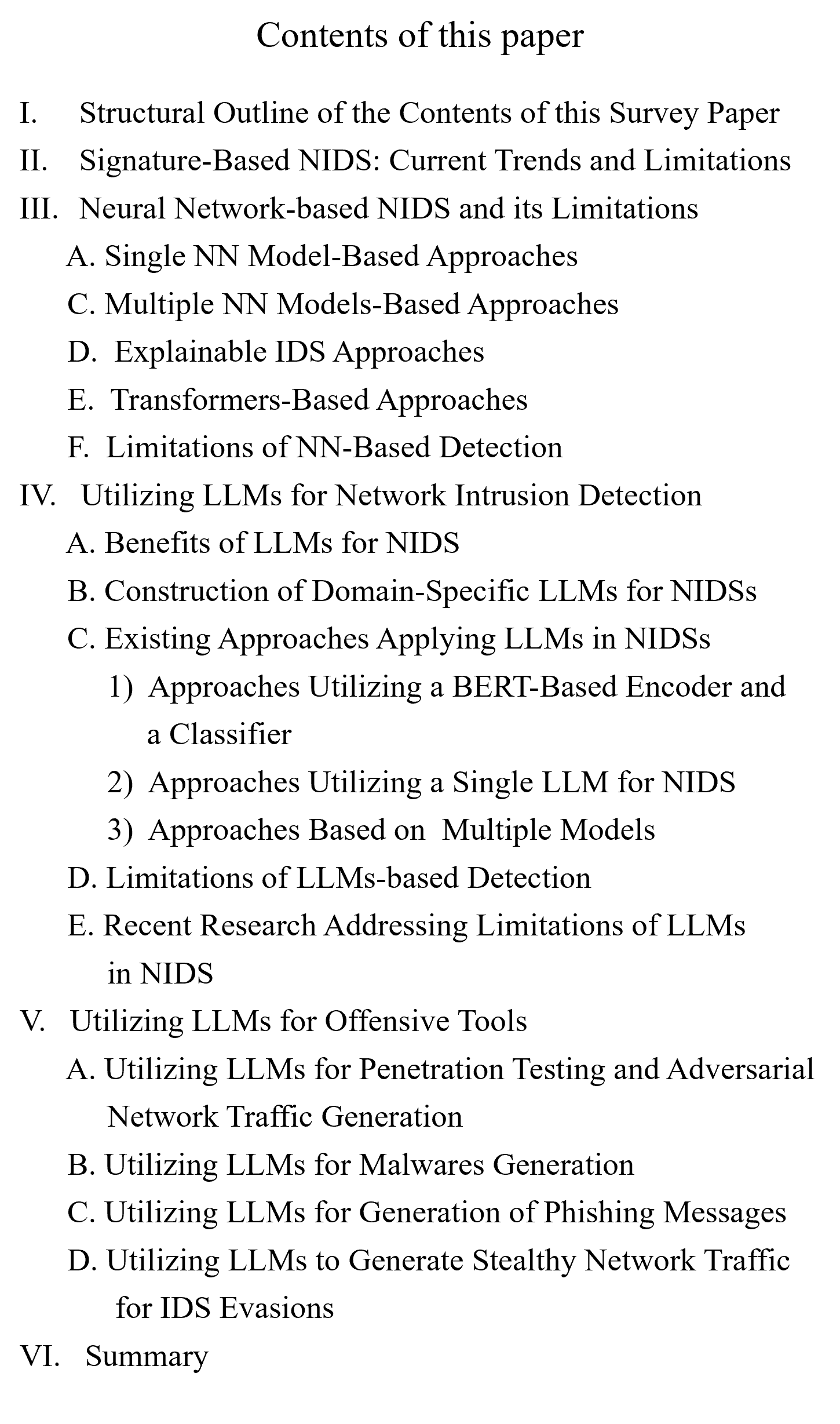}
  \caption{Organization of this survey paper.}
  \label{Contents}
\end{figure}

\begin{table*}[tb]
\caption{Comparison of NIDS Software Tools}
\centering
\small
\begin{tabular}{|m{1.4cm}|m{1.4cm}|m{1.5cm}|m{2.0cm}|m{1.0cm}|m{3.8cm}|m{3.9cm}|}
\hline
& \textbf{Initial Release} & \textbf{Current Manager} & \textbf{Latest Version/Year} & \textbf{Open Source} & \textbf{Key Features} & \textbf{Current Products $\ \ \  \ \ \ $ (Examples)} \\
\hline
Snort & 1998 & Cisco Systems$^{1)}$ & Snort3 / 2021 & Yes & Real-time traffic analysis and packet logging & Cisco Firepower Series \cite{1.2}, Cisco Secure IPS \cite{1.4} \\
\hline
Suricata & 2010 & OISF$^{2)}$ &  8.0.1 / 2025 & Yes & Multi-threaded  NIDS/IPS \& network security monitoring engine & Security Onion \cite{91.9}, SELKS \cite{1.6}, Wazuh \cite{1.8}, Scirius Security Platform \cite{1.82} \\
\hline
Zeek (formerly Bro) & 1998 (Bro) & Corelight$^{3)}$ \cite{1.88} &  8.0.4/2025 & Yes & Passive network traffic monitoring and analysis framework; event-driven scripting language for custom detection & Security Onion \cite{91.9}, Corelight Sensors \cite{1.84} \\
\hline
\multicolumn{7}{|l|}{1) Cisco Systems, Inc. acquired Sourcefire - the company founded by Snort developer Martin Roesch - in 2013.}\\
\multicolumn{7}{|l|}{2) OISF: The Open Information Security Foundation, a community-run non-profit foundation.}\\
\multicolumn{7}{|l|}{3) Zeek was previously maintained by the International Computer Science Institute (ICSI) \cite{1.86}.}\\
\hline
\end{tabular}
\label{table:NIDS_comparison}
\end{table*}

This survey focuses on existing studies of Network NIDSs (NIDS), tracing its evolution from the earliest signature-based methods to neural network (NN)-based approaches, and finally to the latest developments involving cooperation with large language models (LLMs). The strengths, limitations, and current status of each type of method are clearly and concisely summarized, with references to representative studies. This survey also reviews recent studies that aim to mitigate the challenges of LLMs when utilized in NIDS. Additionally, it examines cases where LLMs have been exploited as offensive tools.
Since the scope of this survey is broad, a structural outline of the contents is provided in Figure \ref{Contents} to help readers grasp the overall organization.

\section{Signature-Based Network Intrusion Detection: Current Trends and Limitations}
Signature-based NIDSs remain one of the most fundamental and widely deployed approaches in network security. They rely on predefined rules or signatures that describe known attack patterns, allowing high-accuracy detection of previously observed threats. One of the most representative tools in this category is Snort  \cite{91} (1999), developed by Martin Roesch in 1998 and currently being maintained by Cisco.
Its role has evolved from being a standalone NIDS tool in the early 2000s to becoming a component within larger, integrated security solutions. As of 2025, it continues to be actively developed and maintained by Cisco Systems. The latest version, Snort 3 released around 2021, is now a standard feature of Cisco Secure Firewall Threat Defense (FTD) and serves as the engine for commercial firewall \cite{91.4}. In other words, Snort remains not only a standalone tool but also a core technology within Cisco’s commercial security offerings.

Two other well-known systems are  Suricata \cite{91.2} and Zeek (formerly Bro) \cite{91.8}. Suricata was developed by OISF (the Open Information Security Foundation) in 2009 and its latest version 8.0.1 was released on September 16, 2025. It offers  advantages such as multithreaded processing, automatic protocol detection, trying to enable higher throughput and scalability for modern high-speed networks. In contrast to the strictly signature-based operation of Snort and Suricata, Zeek was originally developed by Vern Paxson  \cite{91.6} (1999) and its latest version is 8.0.3 \cite{91.8}, released on October 15, 2025. It adopts an event-driven and behavior-based approach, providing richer contextual information beyond simple signature matching. 

In modern large-scale NIDS systems, Snort and Suricata often serve as core signature-based detection engines, while Zeek functions as the analytical backbone, providing contextual visibility and network telemetry. Many enterprise and open-source platforms build upon these tools as foundational components of contemporary network defense. For example, they are orchestrated by frameworks such as Security Onion \cite{91.9} , Wazuh \cite{1.8}, and Corelight Sensors \cite{1.84}, forming a hybrid detection ecosystem. A comparison is shown in Table \ref{table:NIDS_comparison}.

 The primary strength of the signature-based NIDS is their high accuracy and low false-positive rate when dealing with known threats. Although it has had a long history and still used in modern  systems as mentioned above,  many studies have mentioned its limitations \cite{61} (2025), \cite{62} (2022), \cite{64} (2025), \cite{65} (2025), \cite{87} (2015) and \cite{86} (2025).  The main limitations are summarized as follows.

\subsubsection{\textbf{Inability to detect unknown (zero-day) attacks} }
  This is because it tries to find intrusions by simply matching.
\subsubsection{\textbf{Vulnerability to evasion and obfuscation techniques} }
Attackers can evade signature matching by encrypting or encoding payloads; 
\subsubsection{\textbf{High maintenance and operational costs (delayed signature updates)}}
To remain effective, signature databases must be continuously updated and tested using threat intelligence. Delays or errors in updates can lead to security gaps or false detections \cite{86} (2025); 
\subsubsection{\textbf{High risk of false positives and false negatives}}
Signature-based systems may generate false alerts by matching benign traffic with similar patterns (false positives), while failing to detect modified or variant attacks (false negatives); 
\subsubsection{\textbf{Scalability and performance limitations}}
Matching a large number of signatures in real time across large-scale networks imposes a heavy computational burden. In high-throughput environments, this can lead to delays or missed detections;
\subsubsection{\textbf{Difficulty in capturing attack context}}
Attack chains or multi-phase attacks often form attack context. Because signature matching is based on isolated patterns, it is difficult to correlate and detect the full life-cycle of an attack, such as $reconnaissance \to intrusion \to lateral movement $. Each stage may look benign in isolation \cite{87} (2015) and \cite{86} (2025). 

 To overcome these limitations, hybrid and anomaly-based NIDSs incorporating machine learning (ML) and deep learning models have emerged since the mid-2010s, combining the interpretability of rule-based systems with the adaptability of data-driven detection. Nonetheless, signature-based approaches remain a critical foundation for network defense infrastructures due to their transparency, stability, and operational maturity.

\section {Neural Network-based Network Intrusion Detection: Capabilities and Limitations}

Due to the above-mentioned limitations of signature-based NIDSs, NN-based detection has attracted significant research interest, with numerous models proposed over the past two decades. There also have been many survey papers on NN-based NIDS technologies. Here, we provide a brief overview organized by categories such as single-model approaches, ensemble methods, temporal sequential models, and explainable models. 
Table \ref{table:nn_nids_approaches} provides a general overview.

\begin{table*}[htbp]
\caption{Representative Neural Network-Based NIDS Approaches}
\centering
\renewcommand{\arraystretch}{1.2}
\begin{tabular}{|>{\raggedright\arraybackslash}m{3.5cm}|>{\raggedright\arraybackslash}m{4.7cm}|>{\raggedright\arraybackslash}m{8.5cm}|}
\hline
\textbf{Category} & \textbf{Model Type} & \textbf{Representative references \& Years} \\
\hline
\multirow{3}{*}{Single Model Approaches} 
& (Smaller models) MLP, CNN, AE, SAE, RBM, etc.& \cite{3.1} (2001), \cite{3.2} (2002), \cite{3.3} (2004),\cite{4.26} (2015),  \cite{4.29} (2011) \\
\cline{2-3}
& (Deep Learning models) DNN DAE, DBN, CNN, etc. &  \cite{4.25} (2014), \cite{4.28} (2015),\cite{4.98} (2018), \cite{4.9} (2018), \cite{4.5} (2019),  \cite{503} (2025)\\
\cline{2-3}
& (Sequence models) LSTM, RNN, GRU, etc. & \cite{4.30} (2016), \cite{5.1} (2017), \cite{4.8} (2017), \cite{4.1} (2018) \\
\hline
Multiple NN Models-Based Approaches & Sequential, Ensemble (Boosting, Bagging), Two-layer, Trigger-based, Multi-layer, CNN-MLP, CNN+LSTM, CNN+BiLSTM, DNN+RBMs, RBMs+DBNs, AlexNet+SKNet, etc.
& \cite{4.21} (2016), \cite{4.6} (2019), \cite{4.91} (2019), \cite{4.7} (2019),  \cite{4.93} (2019), \cite{4.99} (2020),\cite{505} (2020), \cite{4.92} (2020), \cite{4.925} (2020), \cite{4.97} (2020), \cite{505} (2020), \cite{4.94} (2021), \cite{5.4} (2021), \cite{6.4} (2021), \cite{6.1} (2022), \cite{5.5} (2022), \cite{5.6} (2023), \cite{5.9} (2023), \cite{6.3} (2023), \cite{4.95} (2023), \cite{500} (2023), \cite{6.2} (2024), \cite{5.8} (2024), \cite{4.96} (2024), \cite{501} (2024), \cite{502} (2024), \cite{504} (2025)\\
\hline
Explainable Approaches & X-NIDS, XAI, SHAP-based explanations, etc. & \cite{6.6} (2022), \cite{6.5} (2023), \cite{6.71} (2023), \cite{6.7} (2024), \cite{6.8} (2022) \\
\hline
Transformers-Based Approaches & Transformer encoder-based NIDS & \cite{444} (2024), \cite{5.7} (2024), \cite{445} (2025) \\
\hline
\end{tabular}
\label{table:nn_nids_approaches}
\end{table*}

\subsection{Single NN Model-Based Approaches}
In the early days of using neural network (NN)-based models, researchers attempted to implement NIDS using single NN models, starting with a single MLP (Multi-Layer Perceptron) model. \cite{3.1} (2001) trained a backpropagation neural network (of the MLP type) to detect novel attacks (anomaly detection)—they showed how to train neural networks to generalize to previously unseen attack patterns and discussed the issues of feature design and generalization in NIDSs. \cite{3.2} (2002) empirically compared NN models (MLP/backpropagation) and SVMs (Support Vector Machines) to achieve high accuracy on the KDD/DARPA benchmark; they also investigated reduced feature models (using feature subsets) and reported practical performance trade-offs. This is an early and widely cited IEEE conference paper demonstrating the applicability of MLPs to NIDS benchmarks. \cite{3.3} (2004) implemented a multilayer perceptron (MLP) for multiclass classification of attack types (not just binary abnormal/normal) and compared several hidden layer configurations and regularization settings. The paper represents an early 2000s effort to use a single MLP for multiclass intrusion detection (NIDS) tasks. Several other models have also been used. \cite{4.26} (2015) proposed a stacked autoencoder (SAE) deep network for network traffic identification. Experimental results using a real-world dataset collected from the authors' corporate network show that the proposed scheme performs well. 

Many Deep Learning (DL) models have also been explored.  \cite{4.25} (2014) introduces the deep belief network (DBN) to the field of intrusion detection, and proposed an intrusion identification model based on it. \cite{4.28} (2015) used a deep belief network (DBN) trained with a restricted Boltzmann machine (RBM) to classify traffic data. The study claims that their method performs significantly better than the DBN-SVM method \cite{4.29} (2011), where DBN is used for feature extraction and SVM is used for classification. The method in \cite{4.98} (2018) is based on a DNN and uses four hidden layers for classification. The output layer consists of a fully connected network and a softmax classifier. Rectified linear units are used as the activation function for the hidden layers of this model. \cite{4.9} (2018) proposed an asymmetric deep autoencoder  and a stacked deep autoencoder-based DNN classification model for unsupervised feature learning (FL). Only the encoder portion of the autoencoder (AE) is used, operating in an asymmetric manner, aiming to improve the model's efficiency in terms of computation and processing time. Two asymmetric deep autoencoders, each with three hidden layers, are stacked. RF (random forest) is used for classification. Experimental results show that this method achieves better detection performance than \cite{4.25} (2014) and \cite{4.30} (2016). \cite{4.5} (2019) also proposed a model based on a deep neural network (DNN), specifically consisting of multiple stacked fully connected layers. The goal of the study was to implement a flow-based multi-class classification anomaly NIDS. Oversampling and downsampling techniques were used to improve detection performance for minority classes. \cite{503} (2025) proposes a Residual Network-based CNN (ResNet-CNN) model to enhance the detection and classification of network intrusions. The proposed model combines advanced deep learning techniques for automatic feature extraction and robust classification performance.

Many approaches using sequence models (LSTM, RNN, Transformer) have also been proposed. This is in the context of many attacks unfolding over time (e.g., $scanning \to exploitation \to lateral movement$), and sequence models capture temporal dependencies that static approaches miss. \cite{4.30} (2016) used an LSTM architecture to construct an NIDS model with deep learning methods. \cite{5.2} (2016) was one of the earliest LSTM-based NIDS studies, using KDD99 data to show that packet/flow sequences have temporal dependencies that can be exploited by RNNs (recurrent neural networks). \cite{5.1} (2017) utilized system logs as natural language sequences and used LSTM to detect sequence anomalies. This was a highly influential work that demonstrated the effectiveness of sequence modeling for multi-stage and log-based attack detection. \cite{4.8} (2017) applied RNNs (LSTMs) to packet/flow sequences, demonstrating higher detection accuracy than traditional methods. \cite{4.8} (2017) explored how to model NIDS based on DNNs and proposed a DNN approach for intrusion detection using RNNs. Experimental results showed that this approach outperformed traditional ML classification techniques such as J48, Naive Bayes, MLP, SVM, and Random Forest (RF) in both binary and multi-classification tasks. Furthermore, the study investigated the impact of the number of neurons and different learning rates on model performance. \cite{4.1} (2018) proposed an NIDS architecture consisting of a recurrent neural network (RNN) with a gated recurrent unit (GRU), a multilayer perceptron (MLP), and a softmax module. Experiments showed that the GRU was more suitable as the memory unit of the NIDS than the LSTM, demonstrating its effectiveness as a simplification and improvement. Furthermore, the bidirectional GRU performed better.

\subsection{Multiple NN Models-Based Approaches}

Currently, many detection approaches using multiple models (including sequential and ensemble models) have been proposed. \cite{4.6} (2019) proposed a deep hierarchical network based on CNN and LSTM, where CNN is used to extract spatial features of the flow and LSTM is used to extract temporal features. Finally, the features are fed into a fully connected network to classify the flow. The two networks are trained simultaneously to automatically extract spatial and temporal features of the flow. Experiments show that this method performs better than methods using CNN or LSTM alone. \cite{4.99} (2020) proposed an NIDS that combines convolutional neural networks (CNN) and bidirectional long short-term memory (BiLSTM) models in a deep hierarchical architecture. In this system, CNN is used to extract spatial features, while BiLSTM is used to capture temporal features. The authors claim that the multi-class attack detection performance of this scheme is significantly better than that of using only a single model. \cite{505} (2020)  proposes a hybrid CNN-MLP architecture for analyzing the novel and diversified attacks. An effective feature selection and reduction technique also presents  based on random forest regressor along with the correlation parameter.  
\cite{5.4} (2021) and \cite{5.5} (2022) combine LSTM with CNN or feature preprocessing (e.g., PCA) for multi-class detection. \cite{5.6} (2023) introduces a Transformer NN-based NIDS (TNN-NIDS) designed for IoT networks supporting MQTT. By leveraging the parallel processing capabilities of the Transformer architecture, the model accelerates learning and improves detection of malicious attacks. \cite{5.9} (2023) provides a clear comparison between the Transformer-based NIDS and the RNN/LSTM-based NIDS, demonstrating better learning and training efficiency for long sequences. \cite{5.8} (2024) proposes a method for learning network feature representations and detecting feature interactions in imbalanced data using Transformer-based transfer learning.  \cite{504} (2025 integrates AlexNet’s feature extraction module with MLP and incorporates the SKNet attention mechanism to improve the recognition of minority classes.

Ensemble learning has emerged as a powerful intrusion detection strategy, motivated by the need to capture diverse attack patterns that a single classifier cannot typically effectively model. By combining multiple base learners (whether through bagging, boosting, stacking, or hybrid voting methods), researchers have demonstrated that it can improve the robustness, adaptability, and generalization of NIDS performance on benchmark datasets. One prominent direction is the use of stacked ensembles with feature selection. \cite{4.21} (2016) implemented a DNN-based approach using restricted Boltzmann machines (RBMs) and deep belief networks (DBNs). A single-hidden-layer RBM is used for unsupervised feature reduction, and the resulting weights are passed to another RBM to construct the DBN. The pre-trained weights are then fed into a fine-tuning layer consisting of a logistic regression (LR) classifier and a softmax function for multi-class classification. \cite{4.7} (2019) proposed an adaptive ensemble model that uses multiple base classifiers, such as KNN, DT, RF, and DNN, and selects the best classifier using an adaptive voting algorithm. Experiments show that this approach outperforms each method individually. \cite{4.91} (2019) proposed a two-stage deep neural network (DNN) model based on SAEs and a softmax classifier. The model consists of two decision stages: the first stage classifies network traffic as normal or abnormal, and the results are then used as additional features in the second stage to classify normal traffic and various attack types. This approach enables the model to learn meaningful feature representations from large amounts of unlabeled data and perform automatic classification. 
\cite{500} (2023) proposes a relatively lightweight ensemble approach, based on the use of stacking  ensemble technique.  The stacking model is developed using a two-level classification system: a set of base classifiers (three weak classifiers: Gaussian naive Bayes, decision tree and logistic regression), and a single classifier that combines the results of the base classifiers to determine the final prediction.
\cite{501} (2024) presents a hybrid NIDS that combines supervised and unsupervised learning models through an ensemble stacking model. Three ML algorithms comprising a MLP, a modified self-organizing map, and a decision tree.
\cite{502} (2024)  propose a Deep Learning-based Ensemble Framework which is comprised of two layers: the base learner  composed of three robust models: CNN, LSTM, and GRU, and the meta-learner is a DNN model.

\cite{4.97} (2020) also proposed a multi-layer approach to develop flexible and effective intrusion detection models. This architecture combines an unsupervised stage for multi-channel feature learning with a supervised stage that exploits cross-channel feature dependencies. In the unsupervised stage, two autoencoders are trained on normal and attack streams to reconstruct samples. These reconstructed samples are then used to create an augmented dataset, which is input into a one-dimensional convolutional neural network (1D-CNN). The output is flattened and passed through a fully connected layer before entering a softmax layer for final classification. 
\cite{4.93} (2019), \cite{4.92} (2020), \cite{4.925} (2020), and \cite{4.94} (2021) implement sequential detection methods. That is, by using multiple detection models connected sequentially to obtain the final detection result, their detection performance is better than that of a single model. \cite{6.1} (2022) proposed a hybrid NIDS that combines correlation-based feature selection with a weighted stacked classifier, achieving enhanced multi-class detection performance on the NSL-KDD and CIC-NIDS2018 datasets. \cite{6.4} (2021), \cite{6.1} (2022), \cite{6.3} (2023), \cite{6.2} (2024). These works emphasize the importance of combining feature optimization and ensemble modeling for effective detection. \cite{4.96} (2024) proposed a trigger-based two-stage detection method. For the first time, a Riemannian manifold metric was used as a trigger feature, and a mechanism was proposed to update the trigger threshold based on feedback from the second-stage detection results. Experimental results show that this scheme requires significantly fewer second-stage detection calls than previous trigger-based two-stage detection systems \cite{4.95} (2023).

\subsection{Explainable NIDS Approaches}
Research on explainable NIDS has also attracted considerable attention in recent years. Several studies have aimed to introduce AI-based NIDSs with some degree of explainability. \cite{6.6} (2022) proposed a compact two-stage pipeline that uses high-performance ensembles (e.g., XGBoost) for detection and incorporates explainability to validate and reinforce decisions. \cite{6.5} (2023) proposed an explainable AI-based NIDS for IoT scenarios that combines ML detection with SHAP-based explanations to reveal which flow-level and device-level features drive a given alert. \cite{6.71} (2023) introduces an explainable NIDS (X-NIDS) that enables the system to explain its decisions. \cite{6.7} (2024) proposed E-XAI, an end-to-end evaluation framework that systematically measures the effectiveness of black-box XAI techniques (particularly SHAP and LIME) in explaining the decisions of ML/DL models used in network intrusion detection. A review of related research was presented in \cite{6.8} (2022), which was one of the first comprehensive reviews to define X-NIDS as an independent research field, listing black-box and white-box explainability methods, stakeholder requirements (analysts vs. managers), and evaluation challenges.

\subsection{Transformers-Based Approaches}

\cite{443} (2022) proposes RTIDS, a Transformer encoder-based NIDS that uses positional embeddings and a stacked encoder-decoder structure to reconstruct compact feature representations to handle imbalanced high-dimensional streaming data. They demonstrate strong detection performance on common benchmarks. \cite{242} (2023) proposes a Transformer-based IoT NIDS that uses a self-attention mechanism to learn intrusion behaviors from diverse data in heterogeneous IoT environments. \cite{241} (2023) introduces a Transformer neural network-based NIDS (referred to as TNN-NIDS) for IoT networks supporting MQTT. TNN-NIDS addresses the limitations of imbalanced training data by introducing parallel processing and multi-head attention (MHA) layers into the Transformer architecture, thereby enhancing the learning and detection capabilities of malicious activities. \cite{243} (2023) proposes a multi-Transformer fusion NIDS model designed specifically for the Industrial Internet of Things (IIoT) environment. 

\cite{444} (2024) applied Transformer-based spatiotemporal mechanisms to CAN (Vehicle Controller Network) messages, focusing on detecting anomalies in CAN protocol traffic. \cite{5.7} (2024) proposed a Transformer model-based approach specifically for cloud environments. This approach combines the core principles of NIDS with the inherent attention mechanism of the Transformer architecture, enabling deeper analysis of the relationship between input features and various intrusion types, thereby improving detection accuracy. \cite{445} (2025) proposed a self-supervised approach using contrastive learning between Transformer encoders and raw packet sequences. It aims to better generalize to unseen traffic/zero-day anomalies.

\subsection{Limitations of NN-Based Detection}
As mentioned above, signature-based NIDS have many limitations, while neural network (NN)-based detection methods have received widespread attention. Since over 20 years ago, a large number of NN-based detection models have been proposed. As mentioned above, many traditional neural network models have been tried: basic models such as MLP and Deep MLP; sequence and time series models such as RNN, LSTM, and GRU; spatial feature extraction models such as CNN; anomaly detection models such as AE and Variational AE; deep generative and feature learning models such as DBN (Deep Belief Network) and RDM (Restricted Boltzmann Machine); regularization and stability-oriented models such as self-normalizing neural network (SNN); and hybrid models such as CNN + LSTM or AE + DNN, which can simultaneously capture spatial and temporal features. However, even in today's cybersecurity environment, signature-based NIDS still play a vital role. Indeed, signature-based NIDS are now commonly combined with anomaly detection and ML methods to create hybrid solutions that can adapt to evolving threats (\cite{79} (2010), \cite{37} (2016), \cite{94} (2021), and \cite{61} (2025). This is primarily due to various weaknesses and challenges inherent in neural network-based NIDS, such as the stability of detection performance and lack of interpretability. This subsection will address the key specific challenges facing neural network-based NIDS.

\subsubsection{\textbf{The lack of Labeled Data in the Training Dataset}} 
In practice, providing labeled data is extremely difficult. To label traffic data in training datasets as "attack-like" or "benign," security experts must examine logs, payloads, and system behavior. Even experts struggle to determine whether certain anomalous traffic is an attack or simply anomalous but benign. This process is labor-intensive, time-consuming, and requires specialized knowledge. This means manual labeling is costly. Furthermore, new attack types emerge frequently (zero-day vulnerabilities, new malware, adversarial attacks). Since there are no labeled samples when they first appear, datasets quickly become outdated. Privacy and confidentiality are also significant factors. Real-world network traces often contain sensitive or personal data (emails, financial transactions), and organizations are reluctant to share raw data, limiting the availability of publicly labeled datasets. All of this makes building large, clean, labeled datasets often impractical.

However, the classifier must learn the statistical distribution of benign and attack traffic. If there are insufficient labeled samples, the learned decision boundary will be incomplete or biased, that is, the classifier will not be able to capture the diverse patterns of normal and malicious traffic, resulting in misclassification. More labeled data ensures that the classifier captures real patterns rather than random artifacts \cite{66} (2006), \cite{67} (2016), \cite{68} (2019). In addition, insufficient labeled data in the training dataset can lead to overfitting (high accuracy on the training data but poor performance on unseen traffic), especially when the detection model becomes very large and complex, with a large number of parameters \cite{66} (2006), \cite{67} (2016), \cite{68} (2019).

\subsubsection{\textbf{Data Imbalance in Training Datasets}} 
The datasets used in training are often highly imbalanced between different attack classes and between attack samples and benign samples. Due to the limited number of labeled samples for the minority class, the classifier is biased towards the majority traffic. This reduces the recall rate/true positive rate of the attack. Data imbalance is a common problem, particularly evident in NIDS \cite{63} (2020), \cite{69} (2023), and \cite{70} (2023). The imbalance in sample distribution can cause the model to favor samples from the majority class during training and ignore samples from the minority class \cite{71} (2012) and \cite{68} (2019). Data imbalance can lead to model bias. Specifically, it makes the model easily biased towards the majority class data during training, making it ineffective in detecting minority class attacks \cite{68} (2019).

Although some oversampling methods (such as SMOTE) have been used to solve this problem, such methods obviously introduce noise (unrealistic data) and have at least the following side effects. Although some alternatives and improvements to SMOTE have emerged, the core challenges still exist.

\begin{itemize}[leftmargin=10pt]
 \item \textbf{Overfitting problem}. This is because oversampling simply appends duplicate data to the original dataset, which can cause multiple instances of certain examples to become "tied," leading to overfitting \cite{73} (2009).
\item \textbf{Class overlap problem}. This is because if the minority and majority classes are close in feature space, oversampling may generate synthetic samples that fall into the majority region \cite{73} (2009).
\item \textbf{Damage to the data distribution of the minority class}, especially when there are different attack subclasses within the minority class. In this case, SMOTE may incorrectly interpolate across subclasses, generating unrealistic attack samples. This can also negatively impact detection performance \cite{74} (2004).

\item \textbf{Poor performance on high-dimensional data}. Features used in network intrusion detection are often high-dimensional, meaning the data in the feature space is very sparse. The distance-based interpolation used in SMOTE can create unrealistic synthetic points in the sparse space, which significantly degrades detection quality \cite{75} (2019).

\end{itemize}

On the other hand, undersampling algorithms can reduce the number of majority class samples, thereby improving the problem of data imbalance, but there are also some negative effects.

\begin{itemize}[leftmargin=10pt]
     
\item \textbf{Loss of potentially useful information}. Randomly removing benign samples may discard informative patterns. This can cause the classifier to miss important variations in normal traffic, leading to more false positives \cite{76} (1997).

\item \textbf{Risk of underfitting}. If a large number of critical samples are removed, the classifier may fail to correctly model the majority class \cite{77} (2003).

\item \textbf{Risk of misrepresenting true traffic}. In fact, artificially balancing the dataset by removing samples may reduce its realism, thereby harming practical deployment \cite{73} (2009).

\end{itemize}

\subsubsection{\textbf{Complexity of Feature Decision}}
The core of network attack detection lies in isolating attacks within the feature space. Therefore, the separability of different data categories in the feature space is crucial, and feature selection has a decisive impact on detection performance. Feature selection in high-dimensional data remains a significant challenge. Ignoring key features significantly impacts detection performance. However, irrelevant or redundant features can also severely degrade model performance. Consequently, considerable work has been devoted to efficient feature selection, and numerous different feature selection methods (filtering, wrapping, and embedding) have been proposed. However, each of these methods has its own advantages and disadvantages, and no single approach guarantees optimal results in all scenarios. Even for experts, choosing the right method remains a challenge. \cite{78} (2024).

The Internet environment has become increasingly complex, especially with large-scale distributed networks and the Internet of Things (IoT). Many complex attacks are often highly dynamic and exhibit unique spatiotemporal characteristics. \cite{79} (2010), \cite{80} (2016), \cite{68} (2019), and \cite{81} (2020). That is, they evolve rapidly (dynamic), often propagate across hosts or subnets (spatial), and often unfold over time (multi-step, multi-stage, slow-to-slow behavior). This makes it difficult for static ML models to adapt and capture the spatiotemporal characteristics of real-world attacks, thus limiting the effectiveness of ML.

The complexity and dynamic nature of the spatiotemporal characteristics of network attack traffic make it difficult to build models that effectively detect attacks. Specifically, the complexity of the data includes the following aspects \cite{8} (2025):

\begin{itemize}[leftmargin=10pt]
\item \textbf{High dimensionality of network traffic}. It typically contains network layer information, source and destination information, and time series data. Furthermore, it requires interactive consideration, making it more difficult to extract spatiotemporal patterns;

\item \textbf{Heterogeneous network environment}. Especially in the IoT environment, different subnetworks, nodes, and devices make it difficult to determine a set of features that are suitable for attack detection in different network environments \cite{82} (2021);

\item \textbf{Temporal dependency}: Many attack behaviors are temporally dependent, meaning that malicious activities often unfold as a sequence of related events rather than isolated anomalies \cite{68} (2019). For example, reconnaissance, exploitation, and data exfiltration often proceed in stages over time, and their detection requires capturing these sequential patterns. Previous research has shown that recurrent models such as LSTM are effective for intrusion detection precisely because they can learn temporal correlations in attack traffic \cite{84} (2017) and \cite{85} (2016). Similarly, system log analysis frameworks like DeepLog have also highlighted that anomaly detection performance improves significantly when temporal structure is explicitly modeled \cite{83}(2017). These findings suggest that accounting for temporal dependencies is crucial for accurately detecting and classifying evolving cyberattacks. Consequently, effectively detecting such multi-stage, complex attacks is practically difficult and, in many practical cases, nearly impossible.

\end{itemize}

\subsubsection{\textbf{Robustness of Detection Models}}
Robustness refers to a model's ability to maintain good performance as conditions change from training to deployment. This includes both unintentional changes, such as distribution shifts (different network environments, new traffic patterns, new attack variants), and deliberate attempts to evade the model (adversarial attacks)\cite{79} (2010) and \cite{88} (2023). That is, the performance of a detection model can be unstable (unrobust) in two main cases: 1) An NN-based NIDS that appears to perform very well in lab evaluations may miss real attacks or trigger many false positives in production\cite{88} (2023). 2) Deliberate adversarial attacks aim to intentionally cause misclassifications in the detection model. An attacker can craft minimal feature changes that push the input beyond the learned decision boundary without changing the observable malicious intent (e.g., packet timing adjustments, small payload changes, or flow-level feature perturbations). ML models, especially deep networks, learn complex boundaries that adversaries can exploit\cite{89} (2023). The robustness issue has led to the aforementioned current situation: feature-based detection remains the mainstream in network intrusion detection, while the practical application of neural network-based models is very limited, despite the former's many weaknesses mentioned above.

In NIDS research, robustness encompasses the ability to generalize under data or environmental changes and the ability to resist adversarial attacks. Addressing both aspects is crucial for the effectiveness of any ML-based intrusion detector. Although many studies have attempted to address the adversarial attack problem (\cite{90} (2024)), no fundamental solution has yet been found. Furthermore, the first type of unintentional changes, such as distribution shift, stems from problems in the training data, such as data imbalance and a shortage of labeled samples. Therefore, this remains a challenging problem.

\subsubsection{\textbf{Non-interpretability of Detection Models}}
NN models, especially DNN models, often operate like "black boxes," making their decision-making process difficult to understand. Explaining why certain traffic is classified as an attack is crucial for enabling human intervention after an alert is triggered (\cite{63} (2020)). While some research has attempted to mitigate this issue, as discussed in the previous section (\cite{6.5} (2023), \cite{6.71} (2023), and \cite{6.7} (2024), XAI tools have limited adoption among incident responders and struggle to meet the decision-making needs of analysts and model maintainers (\cite{150} (2023).

While NN-based NIDSs offer adaptability compared to signature-based systems, they still suffer from data scarcity, robustness, and interpretability issues. The next section introduces LLMs as a promising complement, explaining how their contextual reasoning and generative capabilities can address these gaps and enable more adaptive intrusion detection.

\section{Utilizing LLMs for Network Intrusion Detection: Capabilities and Limitations}

Given the persistent limitations of NN-based approaches, researchers have explored LLMs to overcome feature engineering bottlenecks and improve explainability. We now examine the advantages of LLMs for NIDS and practical strategies for domain adaptation.

\subsection{Benefits of LLMs for NIDS}
Pretrained language models (PLMs) have demonstrated impressive performance across a variety of natural language processing (NLP) tasks \cite{119} (2023), \cite{220} (2023), \cite{221} (2023), and \cite{222} (2024). Studies have shown that performance tends to improve with increasing model size, especially above certain parameter thresholds \cite{223} (2022) and \cite{221} (2023). Models with large parameter sizes are often referred to as LLMs \cite{222} (2024). LLMs are complex models trained on massive datasets consisting of publicly available text and corpora. They typically contain billions to trillions of parameters. For example, Mistral 7B has 7 billion parameters, considered relatively small, while LLaMA-4 has 2 trillion parameters, making it one of the largest models. Prominent examples include GPT, Claude, Falcon, LLaMA, BloombergGPT, and TigerBot. These models have demonstrated outstanding performance in tasks such as chatbot interaction, text generation, and programming. LLMs can be general or domain-specific. For example, Llama-Fin is tailored for financial applications, supporting tasks such as risk assessment and compliance evaluation through post-training on financial corpora, order tracking data, and preference extraction. Med-PaLM is designed for medical and clinical tasks and fine-tuned using medical literature, case studies, and QA datasets.

The core mechanism behind LLMs is tokenization, which breaks text into smaller units called tokens. LLMs predict the next token based on context, generating coherent and task-specific language. This makes their output both interpretable and effective. LLMs have had a profound societal impact, with applications in areas such as image generation, text composition, and music composition. Their influence spans nearly every field, including cybersecurity.

Traditional ML models, such as rule-based systems or NN-based classifiers, often rely on hand-crafted features and struggle to capture long-range dependencies. To these problems, LLMs offer promising solutions because LLMs can: 1) identify patterns in massive datasets, 2) learn the characteristics of malicious traffic, 3) detect anomalies, 4) characterize the intent behind intrusions, and 5) provide actionable recommendations for security responses.

As the number and complexity of cyber threats continue to grow, the demand for intelligent systems that can automatically detect vulnerabilities, analyze malware, and respond to attacks is becoming increasingly urgent. In recent years, the application of LLM in NIDS has attracted much attention, opening up new avenues for AI-driven network security.

Many studies, including \cite{111} (2024), have evaluated the potential of LLM in NIDS. These studies focus on how LLM can process and understand massive amounts of network log data, autonomously learn, adapt to changing network behavior, and effectively distinguish between normal activities and potential threats. The results show that integrating LLM into NIDS has many practical advantages and can significantly enhance their capabilities:

\subsubsection{\textbf{Continuous Adaptation}}
LLM is highly adaptive, able to learn and update its knowledge as new threats emerge. This ensures that the NIDS can effectively detect new and sophisticated attacks that may bypass traditional rule-based detection mechanisms.

\subsubsection{\textbf{Automated Policy Implementation}} 
LLM enables streamlined automation, enabling complex security policies to be implemented with minimal human intervention. This reduces the likelihood of manual configuration errors and helps prevent misconfigurations that can lead to security vulnerabilities.

\subsubsection{\textbf{Deep Behavioral Insight}}
LLM provides a comprehensive understanding of network traffic behavior, enabling NIDS to identify subtle anomalies that may be difficult to detect. This capability helps identify behaviors that may indicate potential threats or misconfigurations in network devices, allowing proactive preventive measures before an attack occurs.

\subsection{Construction of Domain-Specific LLMs for NIDSs}
Generally speaking, developing specialized LLMs from scratch for network security applications, such as NIDS, is often impractical due to the extensive computational resources required. Fortunately, existing general-purpose LLMs have accumulated rich linguistic and semantic knowledge and exhibit strong generalization capabilities. Instead of building new models from scratch, we can enhance their effectiveness in the network security domain by fine-tuning them using domain-specific datasets. This approach allows us to leverage the rich knowledge embedded in pre-trained LLMs while adapting them to the unique characteristics of network traffic and intrusion patterns.

By combining pre-trained LLMs with target network security data, we can achieve efficient and scalable NIDS implementations that fully utilize existing resources while improving detection accuracy in complex and evolving threat environments. To apply general-purpose LLMs to NIDS, two main approaches are commonly used: continuous pre-training (CPT) and supervised fine-tuning (SFT) \cite{104} (2025).

\subsubsection{\textbf{Continual Pre-Training (CPT)}}
Continuous pre-training involves further training a pre-trained LLM using a large amount of unlabeled domain-specific data, such as \cite{106} (2018), \cite{105} (2020), \cite{107} (2022), \cite{252} (2024), \cite{253} (2024), \cite{254} (2024), \cite{255} (2024), \cite{108} (2024), and \cite{244} (2025). It extends the general pre-training phase of the LLM on a new domain-specific corpus, adapting its language representation to specialized knowledge such as network security logs, network traffic data, or vulnerability reports. Instead of training a complete model from scratch, CPT reuses the general language and reasoning capabilities of a base model (e.g., GPT, BERT, or LLaMA) and applies it to large amounts of network security text or structured event data using the same self-supervised learning objective \cite{105} (2020). This process enables the model to internalize cybersecurity vocabulary (e.g., CVE identifiers, protocol names, log patterns) and improve contextual reasoning capabilities for security-related tasks without losing general language capabilities. CPT can be performed in a variety of ways:

\begin{itemize}[leftmargin=10pt]

\item \textbf{Domain-adaptive pre-training (DAPT)} retrains the model on unlabeled text from a single specialized field (e.g., threat reports or network logs) to shift its distribution toward that domain. 
\item \textbf{Task-adaptive pre-training (TAPT)} continues pretraining on unlabeled data drawn from the specific downstream task (for example, NIDS alert logs) to reduce the gap between pre-training and fine-tuning distributions.
\item More recent work also proposes \textbf{ data-efficient or parameter-efficient CPT}, where only adapter layers or LoRA modules are updated to inject cybersecurity-specific knowledge while keeping the backbone frozen. In intrusion detection, such CPT has been shown to improve anomaly interpretation and threat context extraction from raw IoT or system log data. 

\end{itemize}

\subsubsection{\textbf{Supervised Fine-Tuning (SFT)}} 
Supervised fine-tuning leverages labeled domain-specific data to train a model, directly optimizing its performance on a specific cybersecurity task (\cite{109} (2022),\cite{107} (2022),\cite{256} (2023), and\cite{257} (2023). It aligns a pre-trained or CPT-adapted model with labeled data and a specific objective related to cybersecurity detection, classification, or inference. Compared to CPT's self-supervised adaptation, SFT uses explicit input label pairs, such as "network flow $ \to $  benign/DDoS/port scan" or "log snippet $ \to $ ransomware infection." The fine-tuning objective is typically to minimize the cross-entropy loss of the correct class label or command response, making the model specialized for tasks such as intrusion detection, threat classification, and incident summarization. Because full-parameter fine-tuning updates all parameters of the model, it is computationally expensive, especially for large models. Consequently, parameter-efficient fine-tuning (PEFT) techniques, such as low-rank adaptation (LoRA), have attracted considerable attention. PEFT methods fine-tune only a small number of parameters or introduce additional trainable parameters while keeping the majority of the pre-trained LLM parameters unchanged. This approach significantly reduces computational cost while maintaining performance. Several PEFT techniques have been proposed, including adapter fine-tuning, prefix fine-tuning, hint fine-tuning, LoRA, and QLoRA. These methods offer flexible and efficient alternatives for adapting large models to specific tasks without requiring complete retraining.

\textbf{Adapter tuning} Small neural modules (called adapters) are inserted after the multi-head attention and feed-forward layers of the Transformer architecture. During fine-tuning, only the parameters within these adapters are updated, while the rest of the pre-trained model remains unchanged. This approach significantly reduces computational cost while enabling efficient task-specific adaptation \cite{258} (2021).

\textbf{P-tuning} By introducing trainable cues, optimal cue embeddings for specific tasks are automatically learned. This eliminates the need for manually designed cues and improves performance. This approach can be further enhanced by introducing anchor tags, which help stabilize and guide the learning process, allowing it to be more effectively adapted to specific tasks. \cite{259} (2021) 

\textbf{Prefix tuning} The model parameters are kept fixed, and a small set of continuous, task-specific vectors (called prefixes) are optimized. These prefixes are added to the input sequence and guide the model's behavior during inference, allowing efficient adaptation to specific tasks without modifying the core model. \cite{259} (2021)

\textbf{Prompt tuning} Fine-tuning a language model for a specific task by learning soft hints via backpropagation and using labeled examples to guide the process \cite{260} (2021). LoRA \cite{107} (2022) introduces a small, trainable sub-module in the Transformer architecture. It freezes the pre-trained model weights and inserts a trainable low-rank factorization matrix at each layer, significantly reducing the number of trainable parameters required for downstream tasks. After training, the learned matrix parameters are merged with the original model. QLoRA \cite{261} (2023) builds on LoRA by introducing further optimizations, such as quantization techniques, to reduce memory usage and improve fine-tuning efficiency.

In the field of intrusion detection, SFT is often combined with labeled datasets such as CICIoT2023 or TON-IoT, enabling models to classify new or unseen network events with high accuracy \cite{109.4} (2024) and \cite{109.2} (2025).

In essence, CPT focuses on transferring domain knowledge, while SFT aligns the model with a clear detection goal. CPT enriches the model's internal representation space with cybersecurity semantics, while SFT operationalizes this knowledge to achieve actionable NIDS performance. Modern LLM-based NIDS frameworks often combine the two—first performing domain-adaptive CPT on large amounts of unlabeled security data, followed by SFT on smaller labeled attack datasets—to achieve superior detection of both known and novel threats \cite{109.6} (2024) and \cite{109.8} (2024).

\subsubsection{\textbf{Prompt engineering}}: 

Recent research in natural language processing has highlighted the importance of hint engineering as an emerging fine-tuning approach \cite{264} (2021) and \cite{263.5} (2023). By designing effective hints to guide LLMs toward desired outputs, it can alleviate the training data and resource bottlenecks required for building cybersecurity models. In hint engineering, inserting task-specific hints is particularly beneficial for security tasks involving limited data features. This approach enables LLMs to learn directly from stream-level features in a zero-shot learning manner \cite{265} (2020). In this way, LLMs can extract structured cyber threat intelligence from unstructured data, providing standardized threat descriptions and formalized classifications \cite{266} (2024).

Other emerging techniques also provide valuable insights for building cybersecurity-focused LLMs. Model editing techniques \cite{262} (2023) and \cite{263} (2024) enable direct modification of LLMs to incorporate cybersecurity knowledge without negatively impacting the model's performance in unrelated domains. These methods allow for targeted updates to the model's internal representations, which makes them particularly suitable for adapting LLMs to the evolving threat landscape while retaining general language understanding capabilities.

\subsection{Existing Approaches Applying LLMs to NIDS} 
The topic of applying LLMs to NIDS has attracted widespread attention in the field of network intrusion detection, leading to a substantial body of related research. This subsection introduces representative studies and classify them according to their implementation approaches and application domains. A general overview is presented in Table \ref{tab:llm_nids_summary}.

\begin{table*}[htbp]
\caption{Summary of LLM-based Approaches for NIDS}
\centering
\renewcommand{\arraystretch}{1.2}
\begin{tabular}{|>{\raggedright\arraybackslash}m{2.4cm}|>{\raggedright\arraybackslash}m{8.0cm}|>{\raggedright\arraybackslash}m{2.1cm}|>{\raggedright\arraybackslash}m{3.8cm}|}
\hline
\textbf{Approaches} & \textbf{Main Idea} & \textbf{Applications} & \textbf{Representative references \& Years} \\
\hline
\multirow{3}{*}{\parbox[t]{2.4cm}{\raggedright\arraybackslash Utilizing BERT/Transformer-Based Encoders and a Classifier}}
& \multirow{3}{*}{\parbox[t]{8.0cm}{\raggedright\arraybackslash BERT is primarily used as an encoder — a feature extractor that converts sequences of network events into contextual embeddings, which are then passed to a NN-based classifier for final classification.} }
& Conventional Networks & \cite{114} (2022), \cite{5.8} (2024), \cite{237} (2022), \cite{225} (2023), \cite{228} (2023), \cite{115} (2024), \cite{230} (2025) \\
\cline{3-4}
& & IoT Networks & \cite{112} (2024), \cite{246} (2024), \cite{229} (2024), \cite{227} (2025) \\
\cline{3-4}
& & IoV Networks & \cite{114} (2022), \cite{240} (2024), \cite{239} (2025) \\
\hline
\multirow{2}{*}{\parbox[t]{2.4cm}{\raggedright\arraybackslash Utilizing a Single LLM}}
& \multirow{2}{*}{\parbox[t]{8.0cm}{\raggedright\arraybackslash LLMs and Transformer-based architectures are employed to enhance the performance of NIDSs, with a focus on improving detection capabilities, interpretability, adaptability, and scalability across diverse network environments.}} 
& LLM Used for Detection & \cite{238} (2023), \cite{453} (2024), \cite{232} (2024), \cite{249} (2025), \cite{454} (2025), \cite{244} (2025) \\
\cline{3-4}
& & LLM Used for Interpretation & \cite{113} (2023), \cite{109.8} (2024), \cite{248} (2025) \\
\hline
\multirow{2}{*}{\parbox[t]{2.4cm}{\raggedright\arraybackslash Utilizing Multiple Models}} 
& \multirow{2}{*}{\parbox[t]{8.0cm}{\raggedright\arraybackslash LLM models alone often fail to deliver satisfactory detection performance. And, NN-based NIDSs often struggle to effectively learn from both tabular and textual data. Combining multiple LLMs or integrating LLMs with NN- based models can result in improved overall performance.} }
& Combining Multiple LLMs & \cite{231} (2021), \cite{150} (2023), \cite{441} (2025) \\
\cline{3-4}
& & Combining LLMs and NN Models & \cite{442} (2024), \cite{251} (2024), \cite{151} (2024), \cite{250} (2024) \\
\hline
\end{tabular}
\label{tab:llm_nids_summary}
\end{table*}

\subsubsection{\textbf{Approaches Utilizing a BERT/Transformer-Based Encoder and a Classifier}} 
BERT (Bidirectional Encoder Representations from Transformers) \cite{224} (2019) has demonstrated impressive performance in enhancing various natural language processing (NLP) tasks. In the field of NIDS, BERT is primarily used as an encoder backbone - a feature extractor that converts sequences of network events (e.g., packets, flows, CAN messages) into contextual embeddings(typically [CLS] tokens) . These embeddings are then passed to a NN-based classifier/classification head, such as a multilayer perceptron (MLP), a softmax layer, or an anomaly/attack scoring module, for attack detection. Many studies have explored BERT-based NIDS approaches:

\begin{itemize}[leftmargin=10pt] 

 \item \textbf{Approaches for Conventional Network Environments: }

In \cite{114} (2022), a simple linear softmax layer is used as the classification head. This lightweight design is optimized for embedded deployment, balancing performance and resource constraints; 
\cite{5.8} (2024) utilizes  a combined CNN and LSTM architecture as a classification head for classifying various types of network intrusion attacks. This hybrid approach is capable of capturing both spatial and temporal features, and its effectiveness has been verified by comparison with other deep learning methods. \cite{237} (2022) proposed a scalable multi-anomaly detection model (called AnomalyAdapters) that encodes log sequences using a series of pre-trained Transformers. It uses adapter modules to efficiently learn log structures and anomaly types. This adapter-based approach preserves contextual information, reduces parameter overhead, and can learn across diverse log sources without sacrificing performance. In \cite{225} (2023), network flow sequences are treated as sentences, and a BERT-style encoder is used to model contextual relationships across flows. Compared to traditional ML methods, fine-tuning improves intrusion classification and domain adaptation. 

\cite{228} (2023) proposed a lightweight semantic fusion intrusion detection model. BERT captures semantic features, while BiLSTM learns to fuse features through knowledge distillation, enabling efficient classification.  \cite{115} (2024) proposes a flexible Transformer-based NIDS architecture. This architecture captures long-term network behavior and allows modular replacement of components such as input encoding (including BERT), Transformer layers, and classification heads. The study also analyzed how encoding and classification choices affect performance. \cite{5.8} (2024) used the BERT-based model with a multi-head attention mechanism to mine network data and extract training features, thereby improving the performance of wireless devices. 

 \cite{230} (2025) transforms network traffic into natural language-like sequences, enabling BERT to extract high-quality features. Its bidirectional encoder captures complex contextual information, improving detection accuracy and identifying complex attack patterns. The model efficiently processes sequential data, captures temporal dependencies, and reduces computational complexity, making it suitable for real-time or resource-constrained environments.

\item \textbf{Approaches for IoT Network Environments: }

\cite{112} (2024) proposes a BERT-based architecture for IoT network threat detection, introducing a novel lightweight architecture based on BERT for network threat detection in IoT and Industrial Internet of Things (IIoT) networks. This architecture incorporates privacy-preserving fixed-length encoding (PPFLE) during training, and outperforms traditional ML/deep learning methods.
\cite{246} (2024) performed anomaly-based intrusion detection using a Transformer/BERT-style encoder with BBPE labeling and evaluated the results on an IoT dataset. This is a concrete example of applying LLM technology to IoT intrusion detection. 

\cite{229} (2024) proposes an NIDS using federated learning and LLM to address key constraints of IoT  (edge resources/privacy), while also leveraging Transformer/LLM techniques for network NIDS. BERT is modified to optimize its performance on resource-constrained edge devices. Furthermore, linear quantization is used to compress the model for deployment on edge devices. In other words, the study applies a BERT-style Transformer to the edge/federated environment of IoT/5G and uses quantization/model optimization for edge deployment. 
In \cite{227} (2025), BERT is used for anomaly detection in resource-constrained environments. Its efficient architecture enables low-cost evaluation, making it suitable for IoT applications.

\item \textbf{Approaches for IoV Environments: }

\cite{114} (2022) applied BERT to detect attacks against CAN (Controller Area Network). By treating arbitration NIDS as tokens and using a masked language model objective, BERT can learn periodic protocol sequences and detect multiple attack types in resource-constrained automotive environments. \cite{240}(2024) designs a federated learning-edge-cloud communication architecture for the Internet of Vehicles (IoV) and introduces a feature selection Transformer, FSFormer, combined with a feature attention mechanism to dynamically identify and enhance important features, thereby improving the model's ability to extract key information. In this architecture, mobile users collect and encrypt data, then upload it to edge devices for training. 

\cite{239}(2025) proposes a hybrid network NIDS called IoV-BERT-NIDS, which is designed specifically for on-board and off-board networks in the Internet of Vehicles (IoV) field. The system includes a semantic extractor that converts network traffic data, including CAN packets, into a format suitable for BERT processing. The model is pre-trained and fine-tuned to effectively detect intrusions. BERT plays a key role in capturing bidirectional contextual features, significantly improving the model's generalization and detection accuracy in IoV environments. 

\end{itemize}

\subsubsection{\textbf{Approaches Utilizing a Single LLM for NIDS}} 
Recent research has explored various approaches to enhance the performance of NIDSs using LLMs and Transformer-based architectures. These studies focus on improving anomaly detection capabilities, interpretability, adaptability, and scalability to adapt to various network environments.

\begin{itemize}[leftmargin=10pt]

\item \textbf{LLM used For Detection: }

\cite{238}(2023) proposed a Transformer-based NIDS, CAN-Former NIDS, which predicts abnormal behavior by simultaneously analyzing CAN ID sequences and corresponding message payload values. The model uses fully self-supervised training and token interaction, eliminating the need for manual feature design. 
\cite{453} (2024) proposes a method for detecting cyberattack behaviors by leveraging the combined strengths of LLMs and a synchronized attention mechanism. The effectiveness of the proposed approach is evaluated across diverse datasets, including server logs, financial transaction behaviors, and user comment data. \cite{232} (2024) enhances DDoS detection by converting non-contextual network flows into structured sequences that can be processed by LLMs. The proposal is named DoLLM. DoLLM uses the Llama2-7B model to label flow data and leverages semantic understanding to detect complex attacks such as carpet bombing. This approach significantly improves detection accuracy by capturing subtle flow patterns.  

\cite{249} (2025) proposes an LLM-based detection architecture that comprehensively explains how to use LLMs in network attack detection. The paper explores the three roles of LLMs in pre-training, fine-tuning, and detection: classifier, encoder, and predictor. A DDoS detection case study demonstrated that LLM's contextual mining capabilities excelled in identifying carpet bombing attacks. Together, these studies highlight the versatility of LLM in network security, from improving detection accuracy and explainability to providing adaptive and scalable NIDS solutions for complex and evolving threats. \cite{454} (2025) presents an NIDS approach that leverages a pretrained encoder–decoder LLM (T5), fine-tuned to adapt its classification scheme for attack detection. This anomaly-based method uses statistical features from historical network flows as input. \cite{244} (2025) integrates LLM tokens/embeddings into a dual-path, payload-centric NIDS (token-aware ensemble) to detect payload-level attacks. The focus is on reducing false positives while using LLM embeddings as enriched features for the classifier. It demonstrates how LLM tokenizers/embeddings can transform raw packet payloads into semantically richer representations, improving detection of novel/obfuscated payload attacks. 

% interpretation
\item \textbf{LLM used For Interpretation: }

Although explainable artificial intelligence (XAI) technologies have been introduced to help cybersecurity operations teams better assess AI-generated alerts, their adoption by incident responders has been limited. These tools often fail to meet the decision-making needs of analysts and model maintainers. Meanwhile, LLMs are recognized for their unique approach to addressing these challenges.
\cite{113} (2023) proposed a novel framework (called ChatNIDS) that uses LLMs to explain NIDS alerts in an intuitive language. Designed to assist non-experts, ChatNIDS can interpret alerts and suggest actionable security measures. ChatGPT demonstrated its feasibility, showcasing the potential of conversational AI in cybersecurity. 
\cite{109.8} (2024) uses LLMs as autonomous agents for threat detection and contextual interpretation in IoT networks; LLM reasoning is combined with a network feature extractor for detection and human-readable explanations. It demonstrates the role of LLMs not only in detection but also in supporting operators with interpretable outputs and suggested mitigation steps. 

\cite{248} (2025) introduces eX-NIDS, a framework designed to enhance the explainability of flow-based NIDSs by using LLM. A key component of the framework, Prompt Augmenter, extracts contextual information and cyber threat intelligence (CTI)-related knowledge from flows labeled as malicious by NIDS. This rich, context-specific data is incorporated into the input prompts of the LLM, enabling it to generate detailed explanations and interpretations of why a flow was classified as malicious. The approach is shown to outperform a baseline method, Basic-Prompt Explainer, which does not include contextual information in the LLM prompts. The proposed framework is evaluated using Llama-3 and GPT-4, and the results show that the augmented LLM can generate consistent and informative explanations. These findings suggest that LLM augmented with contextual data can serve as a valuable complementary tool in NIDSs to improve the explainability of malicious flow classification.

\end{itemize}

\subsubsection{\textbf{Approaches Based on  Multiple Models}} 
While Transformers and LLMs show promise for building robust NIDS, LLM models alone often fail to deliver satisfactory detection performance. This limitation stems from the fact that LLMs were originally designed for natural language tasks, not network intrusion detection. Deploying effective NIDS is inherently challenging, especially when trying to accurately identify anomalies in increasingly sophisticated and evasive cyberthreats. Most NIDS research relies on structured features such as network logs, with some exploring text-based features such as payload content. However, traditional ML and deep learning models often struggle to effectively learn from both tabular and textual data. Combining LLMs with NN-based models allows each model to leverage its unique strengths, resulting in higher performance than either model alone. 

\begin{itemize}[leftmargin=10pt]

\item \textbf{Combining multiple LLMs: }

In \cite{231} (2021),  LLMs are used to CAN network security. Since using only forward predictions can miss anomalies that become apparent when checking backward consistency, the authors designed a bidirectional general prediction model (Bi-GPT) that employs two GPT models: one trained on the normal time series (forward GPT) and the other trained on the reversed series (backward GPT). During detection, both models calculate a prediction likelihood for each message. These forward and backward prediction losses are then combined to form a bidirectional likelihood score, which serves as an indicator of whether a given CAN message sequence deviates from normal patterns.  During inference, the bidirectional loss (the average of the forward and backward prediction losses) is used to calculate an anomaly score.  Multi-class classification can be achieved by training with labeled attack patterns, and the output probabilities of the Bi-GPT model can be used as features for downstream classifiers (such as softmax-based or ensemble models). 

\cite{150} (2023) proposes a specialized NIDS called HuntGPT, designed to present detected threats in an easily interpretable form. This system integrates the GPT-3.5 Turbo conversational agent. The results of the study demonstrate that conversational agents based on LLM technology, combined with XAI, can provide a powerful mechanism for generating explainable and actionable insights within the NIDS framework. Through fine-tuning, LLMs can identify patterns in massive datasets and adapt to diverse functional requirements. 
\cite{441} (2025) proposes an intrusion prediction framework for IoT security that is driven by two LLMs: a fine-tuned BART (Bidirectional Autoregressive Transformer) model for network traffic prediction and a fine-tuned BERT model for traffic evaluation. The framework leverages the bidirectional capabilities of BART to identify malicious packets among these predictions.

\item \textbf{Combining LLMs and NN models: }

\cite{442} (2024) proposes a network intrusion prediction framework for IoT security that combines LLMs with LSTM networks. The framework integrates two LLMs in a feedback loop: a fine-tuned Generative Pretrained Transformer (GPT) model for predicting network traffic and a fine-tuned BERT model for evaluating the predicted traffic. An LSTM classifier then identifies malicious packets in the predicted results. 
\cite{251} (2024) discusses the strengths of NN models and LLMs in different tasks and explores how they can be effectively combined to address challenges associated with mobile networks. Specifically, this approach uses synthetic data generated by LLMs to enhance NN-based NIDS. The study proposes the concept of "generative AI-in-the-loop," leveraging the semantic understanding, contextual awareness, and reasoning capabilities of LLMs to assist humans in handling complex or unforeseen situations in mobile communication networks. 

\cite{151} (2024) combines a multilayer perceptron (MLP) and a pre-trained Transformer model, without tokenization, in a neural encoder (CANINE) architecture for processing numerical, categorical, and textual data, respectively. It leverages the advantages of the MLP in capturing complex relationships in numerical and categorical data and the advantages of CANINE's character-based encoding for detailed text analysis.  The experiment results show that combining deep learning and pre-trained Transformer models can significantly improve detection accuracy. 
\cite{250} (2024) proposes an adaptive detection framework, a scalable, real-time NIDS that evolves with emerging threats. It uses a BERT encoder to distinguish between benign and malicious traffic and applies a Gaussian mixture model (GMM) to cluster high-dimensional embeddings. This enables dynamic identification of unknown attack types while maintaining high detection accuracy. 

\end{itemize}

\subsection{Limitations of LLMs-based Detection}
Although LLMs offer clear benefits, integrating them into NIDS introduces critical challenges that affect accuracy, scalability, and trust. As previously mentioned, the complexity of LLMs makes their decision-making processes difficult to interpret. This lack of transparency can lead to trust issues among cybersecurity professionals, who rely on clear explanations to verify the effectiveness of security measures. Another major challenge is the computational cost required to fine-tune LLMs for specific network environments. This process requires significant resources and time, making it impractical for organizations with limited infrastructure. Furthermore, the high cost of model development and deployment can hinder the adoption of LLM-based NIDS solutions, especially in resource-constrained environments. Overfitting is also a concern. When LLMs are fine-tuned on narrow datasets, they may struggle to generalize to unknown threats or network configurations. This reduces their effectiveness in identifying new or evolving attack patterns, which is crucial for maintaining a strong cybersecurity defense.

While LLMs are often valuable, they can sometimes contain incorrect or misleading information.\cite{411} (2023) This section explains the possible causes of these issues and proposes potential solutions to address them in the next section.

\subsubsection{\textbf{Data representation mismatch and domain specificity}} 
Network data (flows, packets, binary payloads, IP addresses, timing information) is very different from natural language. Converting it into tokenized or hinted embeddings so that LLMs can understand it is non-trivial. This mismatch reduces the detection accuracy. \cite{429} (2024) showed that LLMs struggle to accurately detect malicious NetFlow, in part due to these representation issues.

\subsubsection{\textbf{Lack of network intrusion data}} 
Many studies have pointed out the lack of large-scale, high-quality, and up-to-date network security datasets for training or fine-tuning LLMs in the field of intrusion detection. This limits the ability of LLMs to generalize to new threats or real network environments \cite{429}(2024), \cite{430} (2025), and \cite{431}(2025)

\subsubsection{\textbf{High computational cost, latency, and scalability issues}}
LLMs are large in size, require a large amount of inference computation, and are generally memory/GPU-intensive. Real-time LLM inference in high-throughput networks often incurs prohibitive latency unless supported by costly hardware. \cite{429} (2024) notes that while LLMs may help improve interpretability, their "high computational requirements" make them less suitable for pure detection in many deployment environments. While some methods can reduce model size (distillation, quantization, pruning), they may reduce performance or robustness. \cite{432} (2025) shows that optimization/compression increases vulnerability to adversarial attacks or novel attacks.

\subsubsection{\textbf{Adversarial robustness, prompt attacks, and data leakage}}
LLMs are vulnerable to input perturbations (noisy inputs, spelling errors), adversarial examples, and hint-based attacks (hint injection/jailbreaking), which can mislead detection models or lead to false positives/false negatives. These vulnerabilities are particularly concerning in security settings. \cite{433} (2024) showed how carefully crafted hints can force LLMs to produce incorrect outputs even when the semantic input has not changed significantly. Furthermore, research on continuous embedding space attacks (adversarial training in embedding space) shows that LLMs can still be vulnerable even with adversarial defenses\cite{434} (2024).

\subsubsection{\textbf{Explainability, interpretability, and trust}}
Reliable alarm reasoning/explanation is crucial for human analysts to understand the alarms, verify their correctness, and take remedial actions. While LLMs are considered to have great potential to provide some explanation in many cases, they currently struggle with accurate detection, making interpretability more of a supplement than a replacement for traditional methods \cite{429} (2024). Without reliable explanations, there is a risk that false positives and false negatives will be easily accepted, or worse, malicious activity may go unnoticed due to misinterpretation of model confidence.

\subsubsection{\textbf{Evolving threats and model maintenance}}
Cybersecurity is a rapidly evolving field: new attack strategies, zero-day vulnerabilities, protocol changes, and more are emerging. If an LLM is trained or fine-tuned based on historical data, it may not be able to detect or adapt to new attacks. Model drift (schema changes) is a significant problem. \cite{430} (2025) warns that the lack of up-to-date threat knowledge is critical. Furthermore, maintaining an LLM (retraining or continuous learning) is costly and prone to risks (e.g., catastrophic forgetting, introduction of bias).

\subsubsection{\textbf{Overconfidence and misleading}}
Because LLMs can be overconfident or misled (through prompt structure, biased training), false alarms can proliferate, leading to alert fatigue in real operations.

\subsubsection{\textbf{Privacy, legal, and ethical concerns}}
When using LLM on internal logs, payloads, or private network data, there is a risk of exposing sensitive data (e.g., IP, user behavior, credentials) in prompts, logs, or model outputs, which can raise privacy concerns (e.g., data leakage) as well as legal/ethical issues. Data leakage through memorization (where the model remembers parts of the training data) is not trivial (\cite{430} (2025) and \cite{435} (2025). Regulatory compliance (e.g., GDPR: General Data Protection Regulation) may require ensuring that data used for fine-tuning or inference is properly handled; auditing and tracing decision-making processes is even more difficult for large models.

For these reasons, LLMs hold great promise for improving detection, explainability, and automation, but they cannot yet fully replace traditional purpose-built NIDS architectures, especially for real-time detection in high-throughput networks. Addressing these issues requires a combination of domain-specific engineering (feature/token design, fine-tuning, continuous learning), robust evaluation (attack awareness, adversarial testing), and operational considerations (deployment constraints, validation, privacy).

\subsection{Recent Progress in Mitigating LLM Limitations for NIDS}

The following are some specific studies that aim to mitigate the known limitations of using LLMs for NIDS.

\subsubsection{\textbf{Use hybrid architectures: LLMs as augmenting modules, not sole detectors}}
Don't simply replace traditional feature- and statistics-based detectors with LLMs. Instead, combine fast, lightweight front-line detectors (process/rule/anomaly engines) with LLMs for costly tasks such as context analysis, labeling, interpretation, and classification. This reduces latency and cost while leveraging the strengths of LLMs (semantic understanding/interpretation). Recent NIDS surveys and experimental frameworks recommend a hybrid stack as the most practical deployment approach \cite{109.6} (2024) and \cite{436} (2025).

\subsubsection{\textbf{Domain-adapt LLMs via continual pre-training or task-adaptive pre-training before fine-tuning}}
Close the representation gap between natural language pre-training and network data by performing domain-adaptive CPT (continuous pre-training) or TAPT (task-adaptive pre-training) on a corpus of security text, protocol traces, and sanitized payloads before any fine-tuning. LLM-NIDS papers \cite{250} (2024) and \cite{436} (2025) show that CPT can improve token representations of protocol names, CVE NIDS, and log patterns, and reduce phantom reads when interpreting network artifacts. Use a tokenizer and input format tailored to the stream/payload (e.g., hexadecimal/byte-level tokenization plus semantic tags).
 
\subsubsection{\textbf{Parameter-efficient adaptation for frequent updates}}
To adapt models to current conditions without expensive retraining, PEFT (Parameter Efficient Fine-Tuning) methods (LoRA, Adapter, Prefix/Hint Tuning, P-Tuning) can be used. This allows only a small set of parameters to be updated during continuous learning or the infusion of new threat intelligence. This reduces computational effort and the risk of catastrophic forgetting while enabling rapid deployment of updates tailored to new threats. Multiple LLM security studies explicitly recommend using LoRA/Adapter for edge NIDS adaptation, \cite{250} (2024) and \cite{109.6} (2024).

\subsubsection{\textbf{RAG architectures for freshness and transparency}}
RAG (Retrieval-Augmented Generation): Combines LLM with a retrieval store containing the latest threat intelligence, signatures, and network context. Retrieval aligns answers with current indicators, reduces hallucinations, and provides auditability (you can correlate model outputs with retrieved evidence). \cite{436} (2025) and \cite{437} (2025).

\subsubsection{\textbf{Harden inputs: sanitization, canonicalization, and adversarial filtering}}
Treat network inputs as adversarial channels. Before feeding data into the LLM, normalize and sanitize the payload (normalize encoding, remove aliasing artifacts), and run adversarial example detectors or perturbation-resistant preprocessing. Recent LLM security guidelines and NIST guidance emphasize input sanitization and dedicated adversarial filtering modules as effective first-line defenses against evasion and just-in-time injection. \cite{439} (2024) and \cite{438} (2025).

\subsubsection{\textbf{Adversarial training and red-teaming in the development loop}}
Active and persistent adversarial training and red team LLM components using reality evasion techniques (obfuscation, byte-level payload mutation, carefully crafted prompts) are reported to be highly effective \cite{436} (2025) and \cite{438} (2025).

\subsubsection{\textbf{Efficiency engineering: distillation, quantization, and tiered inference}}
Use model distillation and aggressive quantization for low-latency inference; retain a small distilled model for real-time filtering and upgrade to a larger LLM for deep forensics or analyst queries. A paper on deploying Transformers for NIDS claims this layered approach offers the best trade-off between throughput and performance. Careful verification that compression does not reduce adversarial robustness is provided in \cite{109.6} (2024) and \cite{250} (2024).

Understanding these limitations is critical because LLM weaknesses can be exploited offensively. The following section explores how attackers leverage LLMs for penetration testing, attack traffic generation, malware generation, and phishing, and why defensive design must anticipate these threats.

\section{Utilizing LLMs for Offensive Tools}

\begin{table*}[htbp]
\centering
\caption{LLM-based Offensive Capabilities in Cybersecurity}
\begin{tabular}{|>{\raggedright\arraybackslash}m{0.4cm}|>{\raggedright\arraybackslash}m{3.8cm}|>{\raggedright\arraybackslash}m{4.7cm}|>{\raggedright\arraybackslash}m{7.3cm}|}
\hline
\textbf{} & \textbf{Categories} & \textbf{Descriptions} & \textbf{Representative References \& Published Years} \\
\hline
A & Utilizing LLMs for Penetration Testing & Identifying weaknesses in systems, networks, or applications and providing comprehensive reports and recommended improvements & \cite{402} (2023), \cite{403} (2023), \cite{404} (2023), \cite{411} (2023), \cite{412} (2023), \cite{413} (2023), \cite{420} (2023), \cite{452} (2024), \cite{423} (2025), \cite{424} (2025), \cite{425} (2025), \cite{450} (2025), \cite{451} (2025) \\
\hline
B & Utilizing LLMs for Generating Attack Traffic and Malware & Creating malicious network traffic and malicious software, carrying out cyberattacks & \cite{421} (2023), \cite{405} (2023), \cite{409} (2023), \cite{414} (2023), \cite{416} (2023), \cite{415} (2023), \cite{417} (2023), \cite{418} (2023), \cite{419} (2023), \cite{426} (2024), \cite{455} (2025) \\
\hline
C & Utilizing LLMs for Generating Phishing Messages & Creating deceptive messages that imitate legitimate communications, tricking victims into revealing sensitive information & \cite{407} (2018), \cite{406} (2023), \cite{413} (2023), \cite{422} (2023), \cite{410} (2023), \cite{408} (2024) \\
\hline
D & Utilizing LLMs for Generating Malicious Network Traffic Capable of Evading NIDS & Creating malicious traffic in such a way that it is not easily detected by NIDS & \cite{448} (2022), \cite{449} (2024), \cite{426} (2024), \cite{65.5} (2024), \cite{425.5} (2025), \cite{427} (2025), \cite{446} (2025), \cite{447} (2025), \cite{428} (2025), \cite{456} (2025), \cite{457} (2025) \\
\hline
\end{tabular}
\label{tab:llm_offensive_tools}
\end{table*}

The rapid adoption of LLM agents and multi-agent systems has enabled remarkable capabilities in natural language processing and generation. However, they can also be exploited as offensive tools, leading to unprecedented security challenges.
 LLMs have been shown by \cite{401} (2023) to be particularly beneficial during the reconnaissance phase. Using a case study approach, the study explores how LLMs can assist in collecting valuable reconnaissance data, including IP address ranges, domain names, vendor technologies, network topology, SSL/TLS ciphers, ports, and services, and the operating system used by the target. This information helps in the planning phase of a penetration test or network attack, guiding the selection of appropriate strategies, tools, and techniques to uncover risks such as unpatched software and misconfigurations. Penetration testing (or pen testing) is a form of controlled cyberattack used to assess the security of computer systems.  Moreover,LLMs can also be used to generate attack traffic and malware, and even directly carry out attacks\cite{440} (2024). 
 The following subsections provide a comprehensive overview of related work. Table \ref{tab:llm_offensive_tools} presents a general overview of related references.

\subsection{Utilizing LLMs for Penetration Testing}
Penetration testing, typically performed by testers or red teams, aims to identify weaknesses in systems, networks, or applications and provide comprehensive reports and recommended improvements before malicious attackers can exploit them. By simulating real-world intrusions, penetration testing can help organizations discover vulnerabilities and strengthen their defenses against potential threats \cite{400}(2025). Traditional penetration testing is expert-driven and resource-intensive, requiring specialized knowledge, tools, and careful coordination. However, as systems become increasingly complex and the need for frequent testing grows, there is growing interest in automating or enhancing penetration testing with machine assistance. Recent developments have shown that LLMs can support automated penetration testing, offering new possibilities for improving efficiency and scalability.\cite{402} (2023) evaluated the ability of LLMs (specifically Google's Bard and ChatGPT) to generate malicious payloads for penetration testing. The results showed that ChatGPT can generate more targeted and complex payloads, potentially helping attackers build sophisticated exploits. Additionally, \cite{403} (2023) investigated the intersection of LLMs and privilege escalation. The study introduces a fully automated privilege escalation tool designed to benchmark the effectiveness of various LLMs in exploiting vulnerabilities. The results showed that GPT-4 Turbo outperformed GPT-3.5 Turbo and Llama-3 in identifying and exploiting privilege escalation opportunities. Furthermore, \cite{404} (2023) proposes PentestGPT, an automated penetration testing tool based on LLMs. PentestGPT uses LLMs to manage and automate various parts of the penetration testing workflow. It divides tasks into multiple modules (such as reasoning, generation, and parsing) to compensate for the context loss inherent in LLM when handling long or complex tasks. Evaluations show that PentestGPT achieves significantly higher task completion rates than the baseline LLM on real-world targets. Specifically, by combining three self-interacting modules (reasoning, generation, and parsing), the research demonstrates strong performance on a penetration testing benchmark consisting of 13 scenarios and 182 subtasks.

\cite{411} (2023) evaluated the potential of ChatGPT in penetration testing. The GPT model was applied to various stages of penetration testing, including pre-engagement interaction, threat modeling, intelligence gathering, vulnerability analysis, vulnerability exploitation, and post-exploitation. \cite{412} (2023) introduces an LLM guidance tool designed to automate and prototype penetration testing tasks by using prompts to guide the LLM in discovering and exploiting vulnerabilities. Furthermore, challenges such as maintaining focus during testing, handling errors, and handling multi-step exploitation paths were explained. For example, LLM often repeated enumeration commands, resulting in ineffective exploitation of discovered vulnerabilities. \cite{413} (2023) explored how LLM (e.g., GPT-3.5) can assist penetration testers at both a high-level (planning) and a low-level (identifying or executing specific exploits). They implemented a closed-loop system between LLM-generated commands and vulnerable virtual machines via SSH, where the state of the virtual machine is fed back to the LLM to guide subsequent actions. While preliminary, the work demonstrates the feasibility and challenges of enabling LLM to not only plan but also take action (execute) during certain parts of a penetration test. \cite{420} (2023) discusses the use of ChatGPT to generate detailed attack scenarios. It demonstrates the effectiveness of a system that feeds asset management data (OS type, version, device usage, accounts) and vulnerability information published by CISA into ChatGPT, searches for high-threat attack paths, and then outputs attack paths that may be useful for penetration testing and red teaming. 

\cite{452} (2024) introduces an LLM-based agent called HackSynth, designed for autonomous penetration testing. The agent employs a dual-module architecture consisting of a Planner and a Summarizer, enabling iterative command generation and feedback processing. The authors also examine the safety and predictability of the agent’s actions. The work highlights the potential of LLM-based agents in advancing autonomous penetration testing and underscores the importance of robust safeguards.
\cite{423} (2025) presents the VulnBot framework, which uses multiple specialized agents to simulate human-like collaboration during penetration testing. Its architecture splits tasks into multiple phases (reconnaissance, scanning, exploitation), uses a task graph to plan actions, and allows inter-agent communication. Compared to simpler approaches, the framework achieves improved efficiency and accuracy in real-world target experiments. 
\cite{424} (2025) aims to improve the reasoning behavior of LLM agents during penetration testing by using structured attack trees derived from the MITRE ATT-CK framework. By using deterministic task tree constraint reasoning, agents can avoid hallucinatory or wasteful behavior and more efficiently and successfully complete penetration testing subtasks in benchmarks such as HackTheBox. \cite{425} (2025) proposes a multi-agent penetration testing system called xOffense, which uses an open-source LLM (Qwen3-32B) fine-tuned with data from the Thinking Chain. Agents are assigned reconnaissance, scanning, and exploitation phases. The system coordinates these phases and improves performance on benchmarks designed for automated penetration testing. \cite{450} (2025) implements a multi-agent penetration testing framework (called CurriculumPT) that combines curriculum learning with LLM-based agents. The system enables agents to progressively acquire and apply exploitation skills across CVE-based tasks. It uses a structured progression from simple to complex vulnerabilities, allowing agents to build an experience knowledge base. Moreover, it supports generalization to new attack surfaces without fine-tuning, leveraging prior learned strategies. 
\cite{451} (2025) introduces RapidPen, a fully automated penetration testing framework designed to achieve an initial foothold (IP-to-Shell) without human intervention. RapidPen leverages LLMs to autonomously discover and exploit vulnerabilities starting from a single IP address. It integrates advanced ReAct-style task planning with retrieval-augmented exploit knowledge bases and employs a command-generation loop with direct execution feedback. Through this architecture, RapidPen systematically scans services, identifies viable attack vectors, and executes targeted exploits in a fully automated manner.

\subsection{Utilizing LLMs for Generating Attack Traffic and Malware}
Obviously, the technologies reviewed above that are used in penetration testing can also be used to carry out malicious network attacks. 
\cite{421} (2023) explores how criminals are using generative AI to plan and execute ransomware attacks. The research shows that these AI tools significantly lower the barrier to entry for non-technical attackers. The research also found that individuals with IT expertise but lacking other skills can use generative AI to craft more convincing phishing emails. The widespread adoption of generative AI could lead to an increase in the number and sophistication of ransomware attacks. \cite{405} (2023) examines the transformative role of generative AI in social engineering (SE) attacks. The research aims to deepen understanding of the risks associated with this emerging paradigm, its impact on humans, and potential countermeasures. The research demonstrates how AI capabilities in realistic content creation, advanced targeting, and automated attack infrastructure can significantly enhance the effectiveness of these attacks. \cite{409} (2023) also demonstrates the application of generative AI models in SE attacks. These tools can generate highly persuasive, personalized content to enhance the effectiveness of phishing campaigns, including deepfake scams and voice cloning for voice phishing attacks. WormGPT is designed specifically for malicious campaigns, increasing the success rate of business email compromise (BEC) attacks through personalized, convincing emails.

 \cite{426} (2024) introduces AUTOATTACKER, a system that automatically generates attacks using LLMs (LLMs) for complex tasks such as lateral movement, credential acquisition, and other stages of the attack lifecycle. The work focuses on the post-compromise or operational phase of cyberattacks, exploring how LLMs can automate or assist attackers after the initial breach. The proposed framework leverages LLMs for planning, command generation, decision-making, and iterative progression through attack phases. The system simulates or emulates attacker behavior with LLM support. The research demonstrates how LLMs can lower the barrier to entry for sophisticated attacks and provides valuable insights for defense teams to anticipate emerging attack vectors.
 \cite{455} (2025) investigates the use of fine-tuned LLMs for the automated generation of synthetic attacks, including Cross-Site Scripting (XSS), SQL injection, and command injection. A dedicated web application has been developed to enable penetration testers to quickly generate high-quality payloads without requiring in-depth knowledge of artificial intelligence. The fine-tuned model demonstrates the ability to produce synthetic payloads that closely mimic real-world attacks. This approach not only enhances the model’s precision and reliability but also provides a practical resource for cybersecurity professionals to strengthen the security of web applications.

%\subsection{Utilizing LLMs for Malwares Generation}
\cite{414} (2023) discusses the emergence of GPT-based malware and highlights that as traditional malware detection systems evolve to address a wide range of sophisticated attacks, threat actors are now leveraging LLM to develop advanced strategies for infecting new malware. This shift poses significant challenges to traditional detection methods. \cite{416} (2023) demonstrates how publicly available plugins can be combined with LLM, which acts as a proxy between attackers and victims. The study presents a proof-of-concept in which ChatGPT is used to spread malware while evading detection and establishes communication with a command and control (C2) server to receive instructions for interacting with the victim's system. It also outlines the general approach and key elements required to remain undetected and successfully execute an attack. \cite{419} (2023) demonstrates that current large text models can be used by attackers to generate malware. Furthermore, the model's ability to rewrite malware code in various ways is tested. It also highlights the difficulty of GPT-3 in generating complex malware from simple prompts. \cite{415} (2023) demonstrates the powerful capabilities of LLM in generating complex and diverse malware. The ability to automate these processes and increase the sophistication of attacks poses a significant challenge to traditional cybersecurity defenses. It presents 13 examples of cybersecurity-related tasks that ChatGPT can attempt. \cite{417} (2023) also demonstrates the power of LLM in malware creation, where the authors used LLM to create malware such as WannaCry, NotPetya, Ryuk, REvil, and Locky. The paper shows that ChatGPT can generate ransomware code snippets, including encryption procedures and ransom note generation. \cite{418} (2023) investigates the potential for abusing advances in artificial intelligence by using ChatGPT to develop seven malware programs and two attack tools. The authors demonstrate that these models can generate functional malicious code in minutes.

\subsection{Utilizing LLMs for Generating Phishing-Messages}

 \cite{407} (2018) involved using word vector representations of social media posts to train a model to generate spear-phishing messages. \cite{406} (2023) demonstrated the potential of LLM to enhance and scale spear-phishing campaigns. By using GPT-3.5 and GPT-4 to create spear-phishing messages for over 600 UK Members of Parliament, the authors showed that these models significantly lowered the barrier to entry for cybercriminals. 
 \cite{413} (2023) demonstrated that ChatGPT can not only generate malware code and phishing emails but is also highly effective in SQL injection attacks. The authors also demonstrated that AI can generate polymorphic malware and craft convincing phishing emails. 
 
 The article \cite{422} (2023) describes a convincing fake email exchange created using generative AI in which company executives appear to discuss how to cover up financial deficits. With the help of an army of social media bots, the “leaked” information quickly spread, causing the company’s stock price to plummet and causing permanent reputational damage. \cite{410} (2023) is also an article on phishing emails. The paper examined the effectiveness of human-crafted phishing emails compared to those crafted by GPT-3.  \cite{408} (2024) showed that lateral phishing emails generated by LLM were as effective as those crafted by communications professionals, highlighting the key threat posed by LLM in leading phishing campaigns. It revealed that AI-crafted emails, particularly those leveraging internal organizational information, had a high success rate in persuading employees to respond. 

\subsection{Utilizing LLMs for Generating Malicious Network Traffic Capable of Evading NIDS}

Since LLMs have the capability to generate flexible and natural-looking traffic, it is understandable that they are powerful tools for creating attack traffic capable of evading pattern-based NIDS. In fact, they can generate multi-stage attacks, benign-looking malicious traffic, and code that makes detection difficult for traditional NIDS. Moreover, they may also produce noise to degrade the detection performance of NIDS.

\cite{449} (2024) designs an automated system, called EaTVul, to attack DNN-based software vulnerability detection systems.  A surrogate model is first trained based on BiLSTM with an attention mechanism. The averaged attention scores from the attention layer are retrieved to identify the key features that contribute significantly to the prediction. These important features serve  as inputs to ChatGPT, which generates adversarial data. The  generated adversarial data will then be further reviewed and  optimized, and ChatGPT is used again to regenerate the adversarial data. In the above-explained study \cite{426} (2024), LLM acts as a controller that plans and executes attack sequences, integrating with real tools. In \cite{65.5} (2024), the authors introduce that their group generated JavaScript code utilizing LLMs that appears benign but is malicious, in order to bypass NIDS. They produced benign-like  and harder-to-detect code.

Multi-stage attacks can hide their intent and make detection more difficult. \cite{425.5} (2025) investigates whether LLM can act autonomously in multi-stage cyberattacks (reconnaissance, initial access, lateral movement, exfiltration) across multiple hosts. The paper introduces a system called Incalmo, a high-level abstraction layer that allows LLM to specify tasks (e.g., "infect a host," "scan a network"), which are then translated into realistic, low-level commands by agent modules. It also includes services such as environment state tracking and attack graph support. LLM using Incalmo was able to successfully execute multi-stage attacks in 9 out of 10 evaluated network environments. Even the smaller LLM (using Incalmo) was able to successfully execute in 5 out of 10 environments. \cite{427} (2025) provides an overview of the threat capabilities of LLMs when used as autonomous agents in cyberattacks, covering reconnaissance, malicious content/code generation, coordinated or multi-stage attacks, and threats in infrastructure-less mobile or IoT networks. Existing defense approaches and their limitations are also discussed.

Recent studies indicate that LLMs can be integrated into frameworks that automate reconnaissance, exploit selection, fuzzing, and post-exploitation activities—capabilities that generate traffic or behaviors designed to evade ML-based or signature-based NIDS \cite{446} (2025). \cite{446} (2025) proposes an adversarial traffic generation framework, called  AdvTG, to deceive DL-based malicious traffic based on the LLM. And a specialized prompt is designed for traffic generation tasks, where non-functional fields are generated to produce the mutated traffic, while functional fields and target types are supplied as input. This fine-tuning  allows it to generate traffic that remains compliant and functional. Furthermore, reinforcement learning (RL) is utilized to make AdvTG automatically selects traffic fields that exhibit more robust adversarial properties. Experimental
results show that AdvTG achieves over 40\% attack success rate (ASR) across six detection models on six datasets.

LLM-assisted generators can craft payloads that mimic benign application traffic (e.g., HTTP headers, legitimate cookies, or common IoT telemetry patterns) or reformulate exploit code to appear syntactically similar to legitimate samples while preserving semantic validity. Such emulation can defeat detectors that rely on surface-level labeled distributions. Prior work has shown that adversarial examples—small, constrained perturbations to inputs—can cause ML-based NIDS to misclassify malicious traffic as benign \cite{448} (2022). LLMs and generator models can explore the constrained space of protocol-valid perturbations (e.g., minor payload bit/byte changes, timing jitter, header field tweaks) that preserve attack semantics while shifting features outside the detector’s decision boundary. Practical evaluations indicate that such attacks are feasible in many ML-NIDS designs \cite{447} (2025).

\cite{428} (2025) provides an overview of both defensive and offensive uses of LLMs in cybersecurity: detection, red team task automation, and attack assistance (e.g., prompt-based reconnaissance, social engineering). The paper incorporates case studies and practical problems.

\cite{456} (2025) explains that LLM agents can be manipulated to install and execute malware while mimicking normal communication. Techniques like “RAG Backdoor” and “Inter-Agent Trust Exploitation” are used to evade detection. It demonstrate that adversaries can effectively coerce popular LLMs into autonomously installing and executing malware on victims. \cite{457} (2025) shows that LLM-generated noise (text, code, images) can reduce the detection performance and  Code generation tasks were found particularly effective in evading hardware-based detection.

\section{Summary}
This survey traces the evolution of Network Intrusion Detection Systems from traditional signature-based methods to neural network (NN)-based approaches and, more recently, to frameworks incorporating LLMs. Signature-based systems remain foundational due to their transparency and operational maturity but suffer from poor adaptability to zero-day attacks and high maintenance costs. NN-based models introduced powerful learning capabilities and improved anomaly detection but continue to face challenges such as data imbalance, robustness, and interpretability.
LLMs offer promising solutions through contextual reasoning and generative capabilities. These benefits are amplified when combined with domain-specific adaptation strategies, including continual pretraining, supervised fine-tuning, and prompt engineering. Studies on the use of LLMs in both defensive and offensive cybersecurity contexts are outlined. Additionally, the challenges associated with using LLMs for NIDS, along with recent research addressing these challenges, are reviewed."

%Despite their potential, LLMs introduce significant challenges, including high computational overhead, vulnerability to adversarial manipulation, and privacy concerns. Therefore, hybrid architectures that integrate LLMs with lightweight detection engines, retrieval-augmented generation for freshness, and parameter-efficient fine-tuning for adaptability are critical for practical deployment. Ultimately, the convergence of LLMs with traditional NIDS architectures represents a promising path toward scalable, interpretable, and resilient NIDSs capable of addressing the evolving threat landscape.

\bibliographystyle{IEEEtran}  % ← スタイル指定
\bibliography{references}     % ← BibTeXファイル名（.bibは不要）

\end{document}